# Anticipation Next: System-Sensitive Technology Development and Integration in Work Contexts

Sarah Janböcke [1,*] and Susanne Zajitschek [2]

1 SARAH JANBÖCKE Organisational Development and Research GmbH, 45219 Essen Germany
2 OST Eastern Switzerland University of Applied Sciences, Departement Economics, 8640 Rapperswil Switzerland; susanne.zajitschek@ost.ch
* Correspondence: sarah@sarahjanboecke.de

**Abstract.** When discussing future concerns within socio-technical systems in work contexts, we often find descriptions of missed technology development and integration. The experience of technology that fails whilst being integrated is often rooted in dysfunctional epistemological approaches within the research and development process. Thus, ultimately leading to sustainable technology-distrust in work contexts. This is true for organizations that integrate new technologies and for organizations that invent them. Organizations in which we find failed technology development and integrations are, in their very nature, social systems. Nowadays, those complex social systems act within an even more complex environment. This urges the development of new anticipation methods for technology development and integration. Gathering of and dealing with complex information in the described context is what we call Anticipation Next. This explorative work uses existing literature from the adjoining research fields of system theory, organizational theory, and socio-technical research to combine various concepts. We deliberately aim at a networked way of thinking in scientific contexts and thus combine multidisciplinary subject areas in one paper to present an innovative way to deal with multi-faceted problems in a human-centred way. We end with suggesting a conceptual framework that should be used in the very early stages of technology development and integration in work contexts.

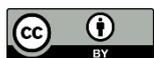

## 1. Introduction

We hit a glass ceiling. Established research, development, and integration methods with or without slight modifications do not necessarily lead to sustainable technological results in work contexts [24,28].

With a look at our VUCA-world, volatility, uncertainty, complexity, and ambiguity became our realities, especially the research, design, and integration of socio-technical systems in work contexts gain increasing critical importance since the rise of Industry 4.0 cyber-physical systems [5,6,9,17]. The introduction of artificial agents, robots, or similar untransparent technological aids in work contexts, to date foremost within manufacturing settings, form a particular form of automation that bear positive and negative effect potential [13,54,59] for the employee and the whole organization. They will change the interaction paradigms within organizations to a whole new extent. The technological transformation of work will proceed and technology will become more proactive, autonomous, and in-transparent to employees [12]. This potential lack of transparency will endanger the understanding work results as one's success [35,58].

Onnasch [40,41] discusses employees' potential competence loss due to monotony effects of automated work settings with various levels of automation [42]. However, when exploring employees' future concerns and their workplace settings, we know relatively less about good human–technology cooperation. The development of current disruptive technology and its enormous speed [1] increases the urgency for new knowledge. It urges us to rethink our general attitude towards technology research, development, and integration [8,36]. We need to consider the complex and paradoxical structures of technological environments discussed within systemic approaches for social systems.

The starting point of our discussion is an observation that many technical systems fail [21] in digital innovation within manufacturing and industry settings because complexity and paradoxes are simply not considered appropriately [29]. For example, with a view on standard prototyping processes (similar to Design Thinking), we try to reduce complexity and uncertainty all too quickly by ignoring the environment or various stakeholders. Triggered by cost reduction, lab situations are far too often the only reality for evaluation. Reality checks remain missing [4].

At the same time, disruptive changes in the labour market are described as the century's most significant challenge ahead [39]. We observe a shortage of skilled workers due to demographic and generational changes that leave us no other choice than to use technological innovations to avoid losing competitiveness in the market [10]. Digitization, artificial intelligence, increased efficiency leave us with the challenge—to master the uncertainty and confusion within the new. However, despite the shortage of skilled workers and substituting technical solutions, the company's core continues to be human [2]. Moreover, this presents us with a further challenge in the jungle of day-to-day operations. Work has a whole range of psychological effects for managers as well as for the workforce. Work contributes to mental health if it leaves room for personal development and competence experience [44]. However, work and its conditions can also contribute to individual reactance and ultimately endanger mental health [7]. This makes it all the more important to develop a secure framework [16] in which innovative work technology will be developed and used/integrated.

This again underlines the demand for a strong attitudinal framework for developing and integrating work technology to prevent increasing health issues of employees due to digitization processes [64] and to decrease economic impacts on technological investments due to employees' reactance [11,47,54].

We see a gap in not covering a networked thinking especially for this specific contexts. Only few articles and methods break away from their specific subject area and look at problems in a multidisciplinary manner. Technological work contexts are in themselves complex systems that are embedded in even more complex organizational structures that find their common ground in the human interface. However, occurring problems are mainly addressed in a mechanistic cause–effect style. Being originally rooted in industrial design and economic science, the authors share the common additional expertise in work and organisational psychology. The authors find their daily challenges in scientific and industrial problems underlining an experience-based motivation to explore new attitudinal frameworks for sustainable technology development in work contexts.

This conceptual work will research existing literature from the adjoining research fields of systems theory, organizational theory, and socio-technical research for our aim to discuss new attitudes and mindsets within the development and integration of socio-technical systems within work contexts. We will conclude by suggesting new directions in thinking within the explored contexts and for future studies.

## 2. Theoretical Background

### 2.1. Understanding the Problem

Organizations are, by their very nature, producers of security who try to eliminate uncertainty from the system with a reduction logic in the form of decisions [61]. Decisions are necessary to convert uncertainty into assumed calculable risk [51]. Socio-technical work systems, which originate in natural science, are foremost developed and integrated with a linear process to understand their users' world. In the name of ensuring efficiency, standard methods of anticipation are aimed at producing controllability and predictability [18]. Linear processes, however, usually lead to path dependency, especially in the case of strategic decisions for technology development and integration [51]. Especially technological work contexts and their organizational interrelations long for a more complex, more social view when aiming for sustainability and the system's health [63].

We deliberately present various subsections to help understanding the research gap, it's environment and the underlying complexity. Since we aim at a new mindset and anticipation methods, we can claim to focus on a so-called work-on-the-system and not within-the-system. This urges for a selection process within the literature review by covering various multidisciplinary facets and later on combining them in an innovative way. We choose to cover the topics of complexity in general, organizational problem-solving logic and common challenges in technology development. We then seek inspiration in system theory and social science with view on organizational structures and socio-technical systems.

2.1.1. Complexity

We want to clarify the concept of (hyper-) complexity [28]. Problems can be distinguished in a continuum between simple, complicated, and complex. Simple problems are characterized by a low number of influencing factors and low dynamics/interaction. Complicated problems are characterized by low dynamics/interactions and many influencing factors. Complex problems are characterized by high dynamics/interaction and many influencing factors [20]. Simple problems can be structured well through intuition or conventional problem-solving approaches. Complicated problems are usually solved with linear cause and effect processes that work well due to the missing dynamics of complicated problems. Complex problems are characterized by many dependencies and links and, in their interconnectedness, show a high dynamic that is difficult to control [20]. Perrow [43] assigns these properties to technical systems and speaks of 'normal accidents' when there are close links and complex interactions [21]. Thus, any form of disruption leads to drastic effects within the system, and similarly, where the unpredictable human being is part of the socio-technological system. This is a close structural coupling with a high level of complexity. System theory understands the human with their psychic system as one system component that keeps the system alive. With a look at socio-technical work systems, we thus can summarise that we are confronted with hyper-complex problems with many interdependencies that each show high dynamics that are rather unpredictable and hard to control. Next, we want to take a look at Gebauer's studies regarding her so-called Logic 1 and Logic 2. Gebauer describes two approaches to tackling complex problems [18].

2.1.2. Logic 1 and Logic 2

Gebauer's Logic 1 and Logic 2 have in common that they deal with the future of VUCA and with uncertainty and complexity. The difference lies in how the future is approached. Logic 1 acts on base of linear-causal thinking. Thus, starting from the past experience with extrapolation attempts to make the future predictable and to anticipate possible risks [18]. Thinking about cause and effect can act as a source of emotional relief. The effects of Logic 1 are the illusion of objectivity and unambiguity. Dilemmas or paradoxes remain unsolved or are attempted to be resolved by decisions based on either-or-thinking. Effects and interactions are neglected [18,31]. Logic 2 is based on systems theory with the aim of navigating the unexpected. Dealing with complex and unpredictable situations is perceived as challenge. The goal is not to maintain protection and security, but rather to deal with currently perceived risks and volatility. While Logic 1 seeks security through repetition, routines and 'patterns of success' [50], Logic 2 is formed by dealing with organizational risks through collective resilience in the form of the right of attitude and mindset. However, having difficulties to endure insecurities during the epistemological research processes is typical as humans. This might be predominantly found in economic organizations since they strive for risk reduction [61]. With ignoring possible, rather complex solutions due to time or even cognitive constraints, we create a development situation similar to a lab development.

2.1.3. Lab-Development

Technology developed in lab situations does not necessarily meet the field's requirements to sustain and survive [3,4]. The difference to a declared lab situation is the assumed field-relevance that nullifies itself by taking the wrong epistemological approach [21,30,60] in the early stages of the development or integration processes—thus producing a lab with a similar development situation. The consequences of paced complexity reduction and false epistemological processes are ignored possibilities that might lead to solutions that sustain a system's health [21,36,63].

In summary, we can say that we observe too many failed technology developments and integrations in work contexts with immense consequences for employees' health and well-being [19]. Reason is often needed to find in the pressure of day-to-day operations that need quick decisions and can produce security [38]. This leads to a rapid reduction in complexity in organizations and their development units when developing and introducing technical systems. This results in enormous economic consequences if the investments fail and enormous human consequences that cannot always be quantified immediately [26,64]. It may sometimes seem that the expectations of the developers and researchers' can superimpose the true knowledge and thus lead to a dysfunctional results [21]. Gransche calls this phenomenon the epistemological accident when he describes a society that knows much about technological risks and dangers but "that is blind to its own new disaster potentials." [21,p.202,translation by the authors].

2.2. Alternative Positions

2.2.1. Social Systems

System theory describes the social system's survival as guaranteed through appropriate communication, i.e., through cooperation [37]. Systems can only be understood if the environment, relevant to their existence, is included within the continuum of possibilities and expectations. Existential environmental spheres can be economy, science, technology, law, politics, society, and/or the public [46].

An organization as a social system is confronted with diverse expectations from the environment, such as customer requests, shareholder demands, union concerns. Subsequently, those expectations lead to irritation, noise in the sense of ideas, opportunities, feedback, and innovation impulses, but not yet to information or even clarity. Irritations lead to the need for decision-making. Attitude and mindset act as a strong (meaning) filter for the process and the behavioural patterns, especially for information acquisition and communication processing [51]. Employees are part of the organization's vital environment from a systemic point of view. Humans and social systems are non-trivial machines [56] characterized by autonomy, laws, or incalculable potential for complexity. Systems theory takes a different look at the world against the background of a constructivist mindset. The question is not 'what is objectively going on', but instead by whom and in what manner the situation is being described and interpreted. This is called a second-order observation [46].

Employees are the observers and observe for the organization. The employees describe, explain, and evaluate their psychological systems and thereby construct reality for the organization in their respective work contexts. This is particularly relevant in dynamic environments because organizations can communicate about it, but they cannot perceive changes in the environment with their senses. Therefore, organizations are dependent on the perceptual abilities of their employees. The organization, with its communications/decisions and the employees' psyche with their awareness/perception are, therefore, two different types of systems. Both represent the existential environment for each other. The cognitive systems of the employees are, therefore, of central importance for the processing of complexity in the organization [63]. The employees perceive, form, evaluate, and test hypotheses and, thus, provide new technological integrations in companies. Companies with external diversity must have adequate internal diversity. This is the only way to ensure the survival or the ability of the system to respond to the environment [29].

2.2.2. Sensemaking and Change

The reorganization and technological development of a social system can mean breaking patterns, and it is always associated with breaking up stability, existing patterns of thought, communication, decision-making, and action, thus causing instability. The power of habit is strong [15]. The fast, involuntary, dominates the slow thinking, the arbitrary in humans [25]. Systems have a persistence tendency that leads back to the old stability, to old behaviour patterns/routines. This is true for organizations which integrate new technologies as the for organizations that invent them. Research processes that want to overcome this persistence tendency by aiming at new, creative solutions or pattern changes require a constant balancing of stability and instability, vigilance, focus, determination, energy, willpower, and perseverance and sensemaking-ability.

Recent articles in experience design, positive design, and design for wellbeing argue that technology should be designed in a way that actively contributes to meaningful, fulfilling work in specific fields [e.g.22,23,32,55]. Few recent, not very common methods like Design Fiction and Techno Mimesis [14,33,34] steer towards a different direction in technology development with the aim to put sensemaking at the focus. Thus, trying to overcome paced conclusions and solutions in designing technological artifacts by leaving the judging observer role while developing new artifacts and strategies. Techno Mimesis, for example, creates a sort of role-play for technology developers or experts with low-key physical artifacts that are used by the participants to build their requisites or stage. By experiencing a technological artifact's role, the subsequently developed technological intervention is most likely different from the original imagined one. Those methods create a forced cut by putting the judge/designer into the participating role. However, a meta-level proposal towards a humanised and social attitude in technological workplace design and development is missing to date, even though those techniques bare the potential to transport new mindsets and approaches in established or even outdated organization structures.

2.2.3. Constructivism

Sense that is created through communication between the system's components is also described in constructivist positions [57]. Large parts of decisions are made unconsciously, and the decisions themselves are mostly based on if-then logic. In constructivism, selection means that something is described, but can also be completely different, which is called contingent information. If we have to choose and decide, we could always do it very differently. However, this has nothing to do with arbitrariness but rather openness [37]. The continuous increase in information, opportunities or expectations of the stakeholders must, therefore, continuously be meaningfully interpreted in communications. In contrast to trivial machines [56] in which output is evaluated deterministically, communicational control attempts in social systems remain contextually complex.

An example: In one of his lectures, Watzlawick speaks of a horse race course with a corresponding visitors' gallery [62]. He describes how the grandstand's railing height has to be increased because people fall backwards over the railing. Looking at it with if-then logic, one could assume that the people who jump over the railing may have a death wish that they live out if their horse bet is lost. From a systemic point of view, however, one quickly comes to the holistic conclusion that communication-cultural contexts were ne-glected in the gallery's construction if a uniform need for distance in communicative situations was assumed. Means: When a high cultural need for distance meets conversation partners who like to be very close, one inevitably moves further and further away and accidentally falls over the railing—two approaches to contextual meaning and two blatantly different results.

2.2.4. Systemic Structure Constellation

Sparrer and Varga von Kibéd [53] and Sparrer [52] describe the systemic structure constellation. It has been used successfully for almost 30 years in organizational, team or market development projects, i.e., in a management context [45]. The systemic structure constellation has its origin in systemic family therapy of Virginia Satir [48] and is used in psychological contexts to illuminate and ultimately heal conflict-laden family systems. Relationships and psychological processes can be mapped spatially with constellation work, which considerably facilitates people's perception and allows all senses to be worked with. It is about the (cyclical) description, explanation, and evaluation in the context of a complex problem. People and their embodiment take over e.g., the role of the organization, the individual employee, the user, the critic, or the role of the innovation itself. These representatives feel and perceive for these systems and generate information with distinctions. It is crucial that the system associated with the problem is provided and not just individual parts.

In practice, however, individual problem areas can also be viewed more closely during constellation work by zooming in, and targeted interventions can be set through circular questions that can set a stuck situation or overly one-sided perspective in motion again. Images can be moved ad-hoc, which means that new information can be created at lightning speed. The aim is to develop a solution in the sense of social rapid prototyping. It is important that it is not about the representatives' individual personality, but about their roles. In this way, many discrepancies, potential problems or patterns of failure as well as successful patterns and solution strategies can be developed step by step realistically and quickly. This would not be possible under pure classic laboratory testing conditions. Similar methods such as the Empathy Map, Persona, Bodystorming or Design Fiction are already used in the design sciences, but mostly remain at the level of individual experiences. In addition, in the concrete application of (social) rapid prototyping.

**3. Concept**

Since work organizations form a special form of social system in which disruptive technology is supposed to be integrated, we draw the connection between the described approaches (Figure 1).

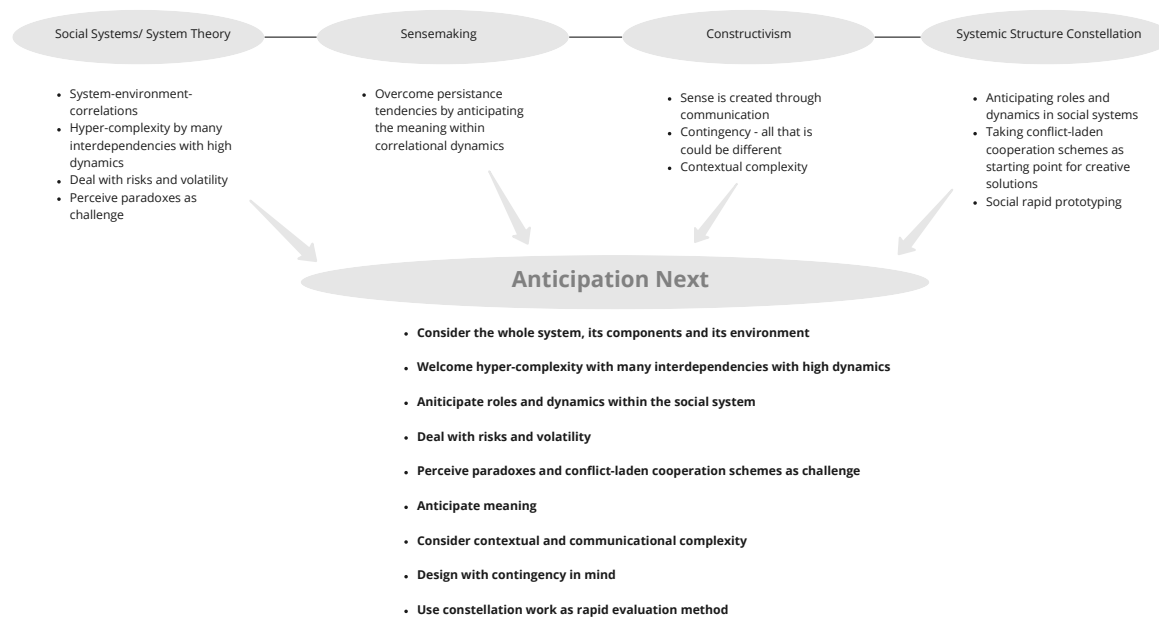

**Figure 1.** Anticipation Next Conceptual Scheme, authors' design.

When developing and integrating new technological processes that critically redefine the whole social cooperative system, we cannot work with linear complexity reduction [51,61] but have to focus on how to successfully deal with hyper-complexity [18,28].

A solution might be to welcome insecurities. Especially the fast-changing work conditions form a problematic field due to their volatility. Thus, the solution must be not the absorption of insecurities but the bearing of and dealing with paradox contexts. The starting point is knowing who I am, what I know (available resources) and whom I know (relationships and interaction with other people). These resources are then related to the potential of the situation [51].

We need pause [49], reflection, and a change of perspective so that the various possibilities are recognized and become information. This is not a 'one-off', but a dynamic and cyclical process. The more laps that are made and the more stakeholders and needs are brought on board, and the more collaborations are entered into, the more stable the project becomes. With this process-understanding, uncertainties will receive enhanced enduring capabilities. The selection of information, the formation of hypotheses, decision-making in the form of interventions, and the action itself are controlled by sense.

We have seen that information is obtained differently from a basic systemic attitude than with a linear-causal mindset. There is no such thing as objective reality. Realities are subjective and can be understood very differently, depending on which perspective you choose—a plea for constructivism.

In the sense of Jan Kruse, who argued for a general attitude towards qualitative research methods rather than directly jumping into detailed analytical steps, we see our task to firstly discuss a general attitude within technological developments before applying detailed research methods [27]
.
However, we see the need to propose and further research a suitable method to transfer the introduced ideas. Coming from Figure 1, we attempt to sketch steps towards a method approach (list below and Figure 2).

(1) Introducing the systemic perspective with lightweight literature and digital workshops. Use Anticipation 2.0 conceptual framework (Figure 1) to get into the idea.
(2) Equally conducting all relevant information about system's components such as:

- Users, users' needs and according behavioural information
- Stakeholder, stakeholder needs and according cooperation dynamics
- Technological components and according subsystems
- Cooperation behaviour information among system's components
- System components' impact factors
- Development unit like scientific researchers or R&D crew of technology providers, development's needs/ intentions, and third-party requirements.

(3) Constructing a system network with equally assigned roles within the cooperation scheme. This means to give each component an equal role but with different impact factors within their cooperation behaviour.

(4) Designing a solution according to the specific system's needs and specifications.

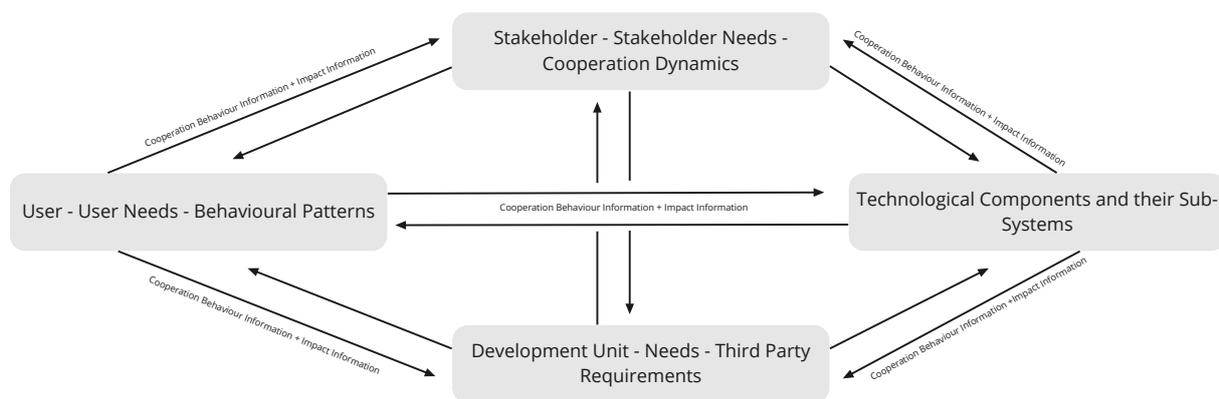

**Figure 2.** Anticipation Next Conceptual Mapping, authors' design.

We argue that transferring the thoughts on system theory we discussed above to a potential technological project in work contexts, i.e., development and integration, will lead to more sustainable and system-healthy solutions. By assigning equal roles to each system component, the likelihood of designing a prosperous and sustainable solution on the market/reality is more likely that when designing around a single component, like focusing only on technological possibility or user needs. Similar to systemic structure constellation, there should be a visible dynamic connection between the system components. That way, the motion of one component might bring the whole system to move in various directions that cannot be anticipated with common methods to date. The discussed positions of dealing with complexity, sensemaking and constructivism thus form base assumptions to prepare for the production of possibly overwhelming complexity. When applying the proposed framework as the basic mindset for technological developments and integrations, the anticipatory ability demanded will most likely increase. Technological research and development to date often neglects the application perspective. On the other hand, organizational development often neglects the user. We thus propose the suggested framework for research and development in science and industry and as well for all those that in later stages will integrate the developed technologies within an organization.

## 4. Discussion

We argue for a system-sensitive processes in the development and integration of socio-technical systems in work contexts, a relaxed mindset towards paradox epistemological research processes, and to break the rules in research, development and integration based on the literature we researched.

The presented positions are rooted in rather diverse fields of expertise. Bringing them together might seem challenging. The impression of too wide-spread concepts might arise while browsing through the many ideas introduced. However, we see a strong connection between the researched positions. Applied to the complex field of technology development in work contexts we strongly believe that the concepts complement each other if summarized in one framework. However, we feel the need to further investigate a concrete method to transport our ideas to avoid concept overload and missing red lines within the communication. Especially for those people that are not familiar with the concepts of system-theory, the claim to introduce themselves to the ideas might be overwhelming. However, since we strongly believe in the value of transferring the presented ideas to technology development, we will deepen our work in the future by presenting methods that ensure proper communication throughout the whole development system.

This work does not aim at presenting validated empirical data. Thus, the shown framework is part of building the base for the development of a new mindset. The framework is ground for building a new normative with the potential to influence new methods, which can then be validated. We aim at future investigations for conducting data and respective validation.

This paper is hardly comparable to common empirical pieces. We thrive to present an explorative paper that combines a literature review with the extraction of a framework. Since we, ourselves, are multidisciplinary, our work is as well. We are aware that our approach bears the risk of causing reactance as it also criticizes well-worn processes. We kindly invite the reader to openly consider the introduced widespread ideas and take first shallow steps within the Anticipation Next mindset. This paper presents the beginning of a research journey and acts as basis for many further discussions.

The issue of introducing a renewed mindset to technology development and integration in work context seems of utmost importance. Referring to the knowledge regarding rising burnout symptoms and digitization, and the huge number of failed investments, we assume a win for many stakeholders, including economic interests, human representatives and technology development units that want their inventions to succeed.

**Author Contributions:** Data curation, S.J.; formal analysis, S.Z. All authors have read and agreed to the published version of the manuscript.

**Funding:** This research received no external funding.

**Institutional Review Board Statement:** Not applicable.

**Informed Consent Statement:** Not applicable.

**Data Availability Statement:** Not applicable.

**Conflicts of Interest:** There is no conflict of interest to declare.